# Monolithic integration of broadband optical isolators for polarization-diverse silicon photonics

Yan Zhang[1,2]†, Qingyang Du[2]†, Chuangtang Wang[1]†, Takian Fakhrul[2], Shuyuan Liu[1], Longjiang Deng[1], Duanni Huang[3], Paolo Pintus[3], John Bowers[3], Caroline A. Ross[2]*, Juejun Hu[2]* and Lei Bi[1]*

1. National Engineering Research Center of Electromagnetic Radiation Control Materials, University of Electronic Science and Technology of China, Chengdu 610054, People's Republic of China

2. Department of Materials Science and Engineering, Massachusetts Institute of Technology, Cambridge, Massachusetts, 02139, USA

3. Department of Electrical and Computer Engineering, University of California, Santa Barbara (UCSB), CA 93106, USA

†Y. Zhang, Q. Du and C. Wang contributed equally to this work

*Corresponding emails: caross@mit.edu, hujuejun@mit.edu, bilei@uestc.edu.cn

# Abstract

Integrated optical isolators have been a longstanding challenge for photonic integrated circuits (PIC). An ideal integrated optical isolator for PIC should be made by a monolithic process, have a small footprint, exhibit broadband and polarization-diverse operation, and be compatible with multiple materials platforms. Despite significant progress, the optical isolators reported so far do not meet all these requirements. In this article we present monolithically integrated broadband magneto-optical isolators on silicon and silicon nitride (SiN) platforms operating for both TE and TM modes with record high performances, fulfilling all the essential characteristics for PIC applications. In particular, we demonstrate fully-TE broadband isolators by depositing high quality magneto-optical garnet thin films on the sidewalls of Si and SiN waveguides, a critical result for applications in TE-polarized on-chip lasers and amplifiers. This work demonstrates monolithic integration of high performance optical isolators on chip for polarization-diverse silicon photonic systems, enabling new pathways to impart nonreciprocal photonic functionality to a variety of integrated photonic devices.

**Introduction**

Nonreciprocal optical devices are essential for controlling the flow of light in photonic systems. These devices include optical isolators placed at the output of each laser to block back-reflected light and circulators to separate signals traveling in opposite directions. Achieving optical isolation on-chip by breaking optical reciprocity has been a major goal of the integrated photonics community.[1,2] An ideal integrated optical isolator should feature several important characteristics including: monolithic integration, high isolation ratio and low insertion loss, broadband operation, polarization diversity, and multi-material platform compatibility. Achieving these functions in a photonic integrated circuit (PIC) is a critical challenge requiring device design combined with materials development and integration.

Several approaches have been made to achieve isolation, including the use of nonlinear effects[3,4] or active modulation of refractive index[5,6], but passive devices based on magneto-optical (MO) effects are the most attractive solutions. MO devices may be based on mode conversion via the Faraday effect[7,8] as used in bulk isolators, but the birefringence of on-chip waveguides favors devices based instead on a non-reciprocal phase shift (NRPS) including ring resonators, multimode interferometers and Mach-Zehnder interferometers (MZIs)[9-16]. The best-performing MO materials in the near-IR communications band are yttrium iron garnets substituted with Bi or Ce to increase their Faraday rotation[17-20]. Integration of garnet into silicon PICs has been accomplished via wafer bonding[21] and via monolithic integration[18,20].

Considerable progress has been made in both device design and materials development, primarily focused on transverse magnetic (TM) mode devices in which the garnet is placed on the top or bottom surface of the waveguide. Wafer-bonded TM ring resonator (RR) isolators exhibit isolation ratio up to 32 dB and insertion loss as low as 2.3 dB[12,13], but with low isolation bandwidth. MZIs exhibit higher bandwidth, and TM MZI devices have been fabricated on single-crystal garnets[16] or by wafer bonding[14,15] (Table 1). However, on-chip lasers produce transverse electric (TE) light whose isolation requires symmetry breaking transverse to the waveguide[22]. TE isolation has been demonstrated by Faraday rotation[8], by device fabrication on single crystal Ce:YIG[23], and by combination of a TM isolator with mode converters[24-26] but these solutions are large in area, difficult to integrate, lossy due to extra polarization rotators, or require complex fabrication processes.

Here we address all the aforementioned requirements for practical on-chip optical isolation by demonstrating monolithically integrated magneto-optical isolators on silicon and silicon nitride (SiN) waveguides operating for both TE and TM modes with high isolation ratio, low insertion loss, small footprint and broadband optical isolation. We demonstrate the first fully-TE broadband isolator by depositing high quality magneto-optical garnet thin films on the sidewalls of silicon interferometer waveguides, and the multi-material platform compatibility of this technology by demonstrating the first monolithic optical isolator on SiN. Both TM and TE isolators show the best performance to date among broadband optical isolators on silicon, with optical isolation up to 30 dB and insertion loss as low as 5 dB.

## Results

**Device design and fabrication**

Figure 1a illustrates the generic layout of the broadband isolator, which consists of a silicon Mach-Zehnder interferometer (MZI) with serpentine waveguide arms embedded in $SiO_2$ cladding. Window sections were etched into the top $SiO_2$ cladding to expose the silicon waveguide on alternating serpentine segments. A blanket magneto-optical Ce:YIG(100 nm)/YIG(50 nm) film stack was then deposited on top of the device. For the TM isolators, the entire top surface of the Si waveguide within the windows is covered with the MO film (Fig. 1c), whereas for the TE devices the waveguide top surface is masked by SiO$_2$ such that the film only deposits on one side of the waveguide (Fig. 1b). (The NRPS cancels out if the film is deposited on both sides of the waveguide.) When the film is magnetized under a unidirectional magnetic field, nonreciprocal phase shifts of opposite sign are induced in the two interferometer arms, leading to constructive (destructive) interference of forward (backward) propagating waves and optical isolation. The design therefore uniquely features a small footprint, large bandwidth, and compatibility with a simple unidirectional magnetization scheme.

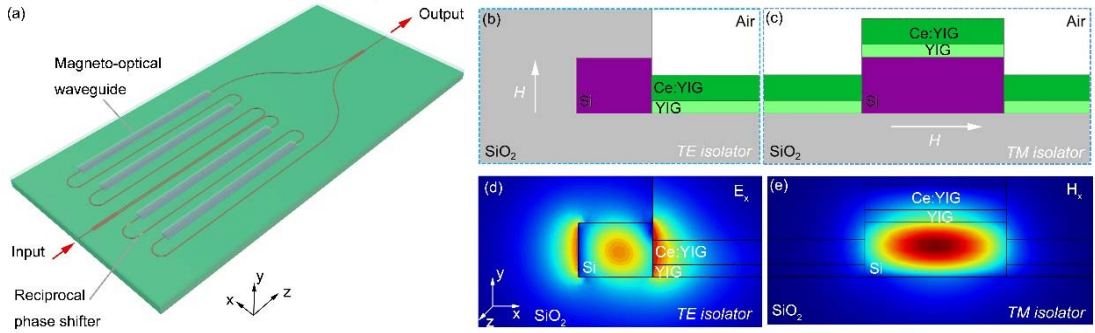

**Fig. 1 Schematics of the TM and TE isolators. a,** Illustration of the device layout **b,** Sketch of the magneto-optical waveguide cross-section for the TE isolator. The magnetic field is applied perpendicular to the film plane. **c,** Sketch of the magneto-optical waveguide cross-section for the TM isolator. The magnetic field is applied in the film plane. **d,** Simulated E$_x$ field distribution of the fundamental TE mode for the magneto-optical waveguide. **e,** Simulated H$_x$ field distribution of the fundamental TM mode for the magneto-optical waveguide.

The simulated modal profile is shown in Fig. 1d and Fig. 1e. NRPS $\Delta\beta$ of the TM and TE modes are given by:

$$\Delta\beta(TM) = \frac{2\beta^{TM}}{\omega\varepsilon_0 N} \iint \frac{\gamma}{n_0^4} H_x \partial_y H_x \, dxdy$$

$$\Delta\beta(TE) = \frac{2\omega\varepsilon_0}{\beta^{TE} N} \iint \gamma E_x \partial_x E_x \, dx dy$$

where $\beta^{TM}$ and $\beta^{TE}$ are the propagation constants for the fundamental TM and TE modes, $\omega$ is the frequency, $\gamma$ is the off diagonal component of the permittivity tensor of the magneto-optical material, $\varepsilon_0$ is the vacuum dielectric constant, $N$ is the power flux along the z direction, $n_0$ is the index of refraction of the magneto-optical material, and $H_x$ and $E_x$ are the electromagnetic fields along the $x$ direction. Considering Faraday rotations of Ce:YIG (-3000 deg/cm) and YIG (220 deg/cm, Supplementary Fig. S2), the simulated NRPS are 16.2 rad/cm and 18.9 rad/cm for TE and TM waveguides respectively, which stipulate nonreciprocal phase shifter waveguide lengths of 968 μm and 830 μm to achieve a total nonreciprocal phase difference of π on both arms. A reciprocal phase shifter (RPS) producing 50.5 π phase shift (16 μm and 22 μm long Si waveguide for TE and TM modes respectively) is introduced in one arm of the MZI devices, creating a total phase difference of 50 π and 51 π for the forward and backward propagating light respectively. The serpentine MZI layout enables a small device footprint of 0.87 mm × 0.34 mm for TE isolators and 0.94 mm × 0.33 mm for TM isolators.

**Performance of TE and TM optical isolators on Si**

Figures 2a and 2c show top-view optical micrographs for both types of isolators. The sections with open $SiO_2$ windows appear darker. For the TE device, the oxide windows are smoothly curved on both ends to allow near-adiabatic mode transformation between waveguide segments with and without garnet with minimal loss. Cross-sectional scanning electron microscope (SEM) images taken within the window sections (Figs. 2b and 2d) indicate that the Ce:YIG/YIG polycrystalline garnet coated waveguides closely follow our designed geometries illustrated in Figs. 1b and 1c. The garnet thin films also exhibit excellent crystallinity and chemical homogeneity up to the Si/MO oxide interface for both devices, evidenced by high resolution tunneling electron microscopy (TEM) and energy dispersive spectroscopy (EDS) analysis presented in Supplementary Fig. S1.

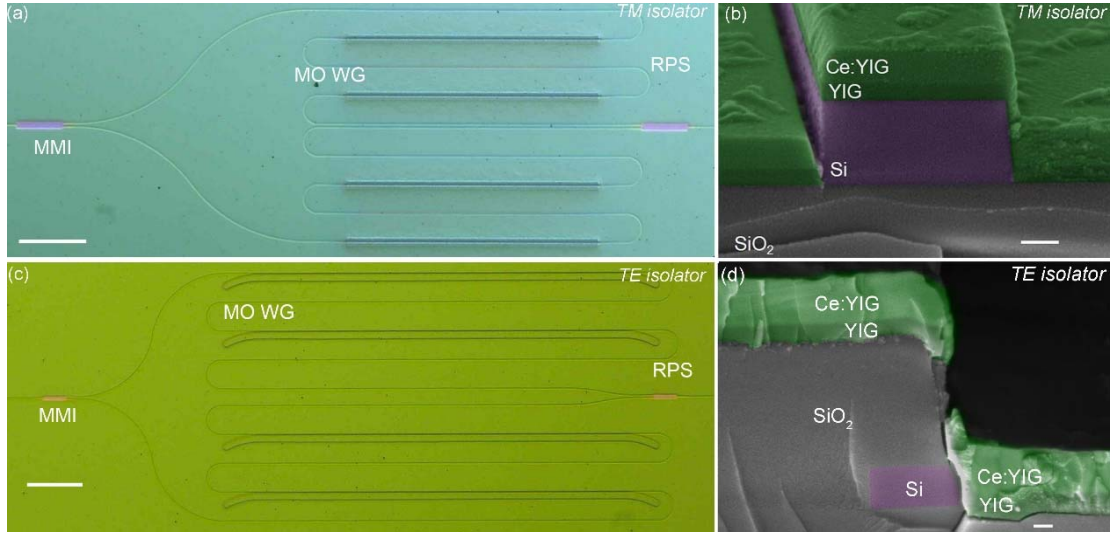

**Fig. 2 Optical microscope and SEM images of the TM and TE isolators. a** and **c**, optical microscope image for the TM and TE isolators respectively. The scale bars are 100 μm. **b** and **d**, cross-sectional SEM image of the magneto-optical waveguides for the TM and TE isolators, respectively. The scale bars are 100 nm. In **b** and **d** the MO layer is colored in green and the Si waveguide in purple.

Figure 3a plots the transmission spectra of the TM mode optical isolator under a uniaxially applied magnetic field of 1000 Oe, together with a reference silicon waveguide on the same chip. The interleaving fringes on the forward (red) and backward (blue) propagating spectra are detuned by approximately half a free spectral range (FSR). The result shows that the device attains a nonreciprocal phase difference of π for the forward and backward propagating modes consistent with our design. Fig. 3c shows the measured (dots) and modeled (lines) isolation ratio and insertion loss around 1574.5 nm wavelength, where the model takes into account waveguide dispersion of the reciprocal and nonreciprocal phase shifters. The maximum isolation reaches 30 dB. The 20-dB and 10-dB isolation bandwidth of this device is 2 nm and 9 nm respectively. The device bandwidth can be readily increased by reducing the RPS waveguide length. Across the entire 10-dB isolation bandwidth, the device shows low insertion loss of 5-6 dB, which represents the lowest insertion loss measured in a broadband on-chip isolator. In a PIC application, the isolator may be biased using integrated permanent magnets to avoid demagnetization by external fields, but the device itself operates at zero applied field and its performance is stable in applied fields up to 150 Oe. Due to reflections when coupling the fiber to the isolator chip, we cannot directly measure the return loss of the isolator. Instead we simulated the return loss of this device by considering the reflection at the MMI junctions and Si/MO waveguide junctions, leading to an upper limit of -24.4 dB, assuring low reflection from the isolator device itself. Further optimization of the Si/MO waveguide junctions may lead to return loss well below -30 dB.

Figure 3b shows the transmission spectrum of the TE mode optical isolator, a maximum isolation ratio of 30 dB, insertion loss of 9 dB and 10 dB isolation bandwidth of 2 nm is achieved at 1588 nm wavelength. To the best of our knowledge, this is the first fully-TE broadband isolator integrated on silicon where no polarization rotators are required. The relatively high insertion loss is due to an insufficient nonreciprocal phase shift in the TE mode MO waveguide, as one can see from the insufficient shift of the forward and backward propagating spectrum. The NRPS of this device, 3.6 rad/cm, is lower than that of the designed value of 14 rad/cm (Supplementary Section 7). The difference is possibly due to a lower magneto-optical effect of the Ce:YIG thin films grown on the silicon waveguide sidewalls, or due to a small air gap between the Si waveguide and MO thin films[25], which may be improved by optimization of the thin film deposition process. The interference fringes in the transmission spectrum of this device are due to Fabry-Pérot interferences from the cleaved waveguide facets, which can be minimized by designing spot size converters or using grating couplers.

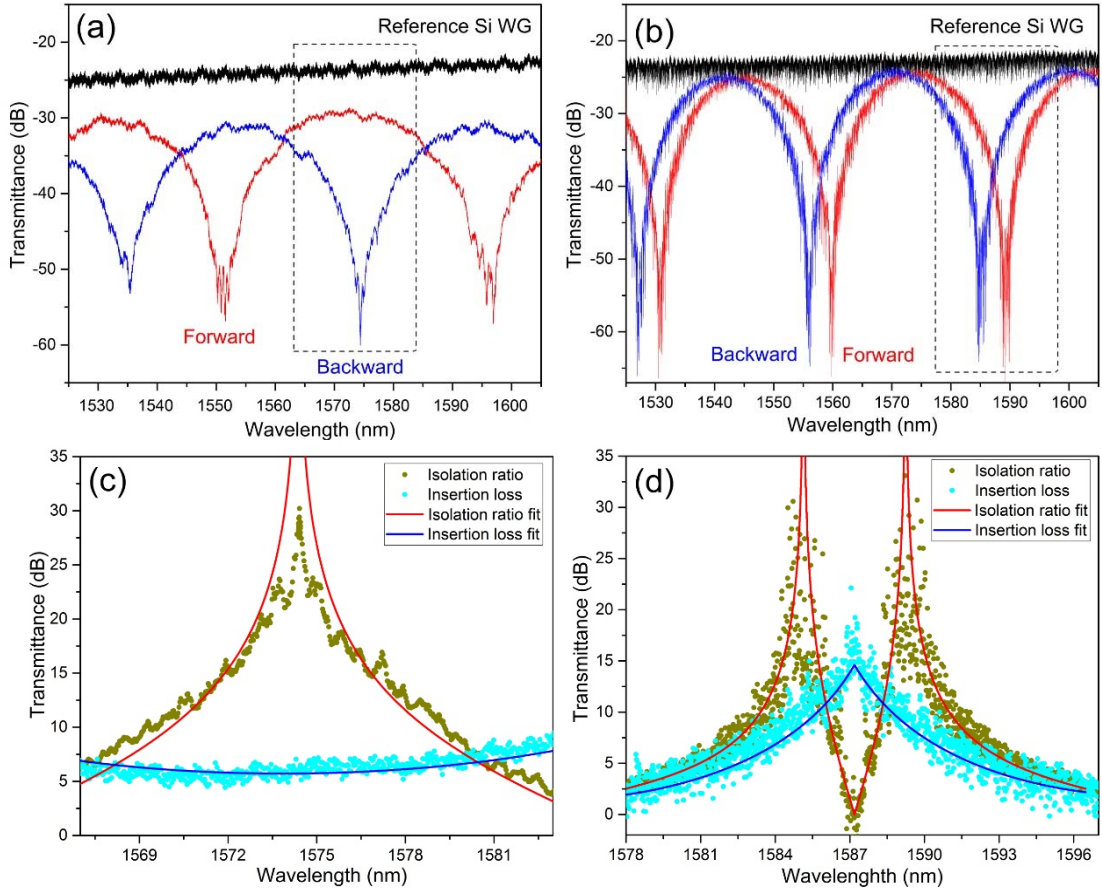

**Fig. 3 Forward and Backward transmission spectra of the isolators. a and b,** Transmission spectra of the TM and TE mode isolator respectively. The corresponding isolation ratio and insertion loss in the dashed regions are shown in **c,** for the TM isolator and **d,** for the TE isolator respectively.

**A monolithic TE optical isolator on SiN**

Besides Si, SiN is another standard waveguide material widely employed in silicon photonics platforms offering unique advantages such as back-end-of-line (BEOL) compatibility and visible light transparency over Si. To date, integrated optical isolators have not yet been demonstrated on the SiN platform[28]. Here we further show that our monolithic approach can be equally applied to isolator integration on SiN through demonstration of the first TE-mode isolator on SiN. The isolator comprises a SiN racetrack resonator encapsulated in $SiO_2$ cladding. The bending radius and the straight region are both 150 μm and the waveguide has a dimension of 400 nm height and 800 nm width. A window is opened in the cladding to expose one waveguide sidewall similar to the Si TE isolator design depicted in Fig. 1b. The fabricated device is shown in Figs. 4a (top-view optical micrograph) and 4b (cross-sectional SEM). It worth noting that unlike the TM resonator isolator design demonstrated previously[10], the window can extend along the entire resonator without cancelling out NRPS as the magnetic field is applied along the out-of-plane direction. In our SiN device, the window covers the resonator device except the coupling section to avoid changing the coupling condition to the bus waveguide.

Transmittance spectra of forward and backward propagation light are displayed in Fig. 4c, which yields an insertion loss of 11.5 dB and an isolation ratio of 20.0 dB at resonance. We repeated the measurement multiple times, and the data in Fig. 4d consistently show a resonant peak shift of (15 ± 2) pm upon reversing the light propagation direction. The result unambiguously validates nonreciprocal light propagation in the SiN device.

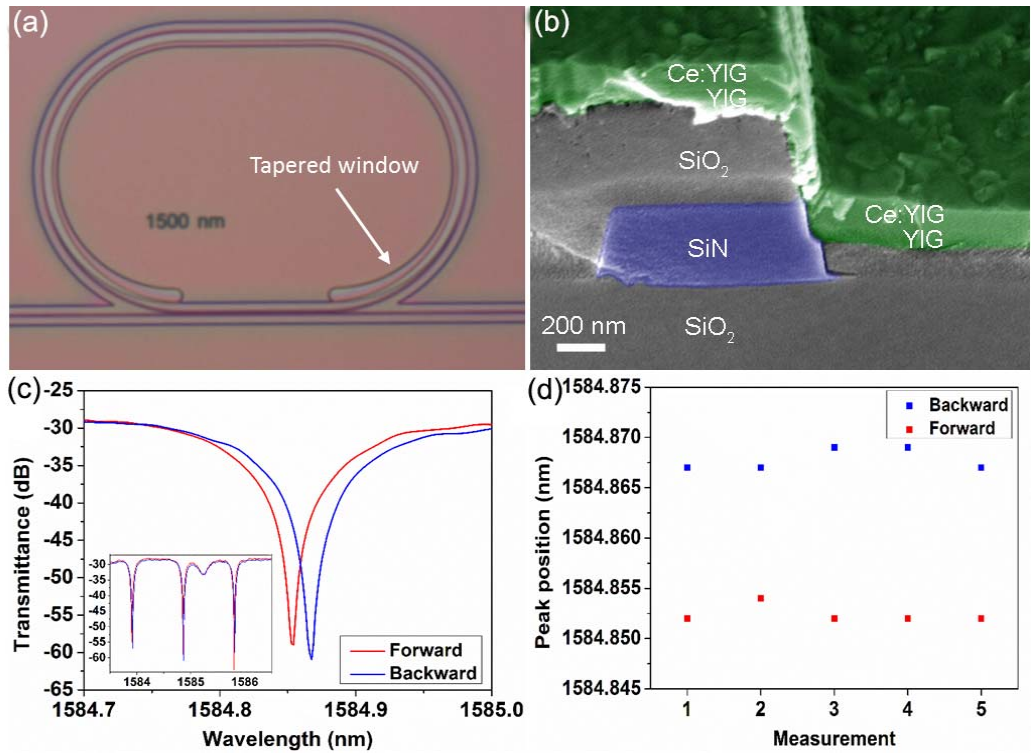

**Fig. 4 A SiN based microring magneto-optical isolator. a,** Optical microscope image of the SiN microring isolator. The gap between the bus waveguide and the racetrack resonator is 1500 nm. **b,** Cross-sectional SEM image of the SiN magneto-optical waveguide. **c,** Forward and backward transmission spectrum of the isolator. The inset shows the transmission spectra of three resonance peaks of the same device. **d,** The peak positions of the forward and backward propagation light for multiple measurements.

The ring resonator device has a narrow bandwidth compared to MZI devices. The resonance frequency can be thermally tuned, and ring devices have been integrated in several applications such as optical communication modules, sensors and frequency combs. We also fabricated a MZI TE SiN device following a similar geometry to that of the Si TE MZI presented in Supplementary Fig. S5, showing that broadband SiN TE mode isolators are also possible. This isolator achieved a maximum isolation ratio of 18 dB, insertion loss of 10 dB, and 10 dB isolation bandwidth of 3 nm at 1581 nm wavelength.

## Discussion

To benchmark the performance of our device, Table 1 compares the device performance of broadband optical isolators on silicon. For TM devices, our device claims a high isolation ratio, the lowest insertion loss, and the smallest footprint. These results demonstrate the possibility to monolithically integrate optical isolators on silicon with performance approaching that of bulk optical isolators[29]. For TE devices, our work demonstrates TE nonreciprocal phase shifters and optical isolators on silicon and SiN for the first time. The ability to deposit high quality polycrystalline garnet thin films both on the top and sidewalls of Si and SiN waveguides is significant because it allows introduction of optical nonreciprocity in planar photonic devices by filling trenches, covering nanostructures or forming photonic crystals, therefore enabling new pathways to impart nonreciprocal photonic functionality to a variety of existing photonic integrated circuits.

**Table 1 Comparison of device performance at 1550 nm for broadband optical isolators on Si**

| Device Type | Isolation ratio (dB) | Insertion loss (dB) | Size (mm x mm) | Monolithic/ Bonding | Polarization | 10 dB Bandwidth(nm) | Ref. |
|---|---|---|---|---|---|---|---|
| Si MZI | 30 | 5 | 0.94 x 0.33 | Monolithic | TM | 9 | This work |
| Si MZI | 30 | 9 | 0.87 x 0.34 | Monolithic | TE | 2 | This work |
| SiN MZI | 18 | 10 | 3.2 x 1.0 | Monolithic | TE | 3 | This work |
| Si MZI | 27 | 13 | 1.5 x 1.5 | Bonding | TM | ~18 | 31 |
| Si MZI | 25 | 8 | 4 x 4 | Bonding | TM | ~1 | 32 |
| Si MZI | 32 | 22 | 4 x 4 | Bonding | TE | ~1 | 25 |
| Si MZI | 30 | 8 | 1.7 × 0.3 | Bonding | TM | ~50 | 14 |
| Si Faraday rotator | 11 | 4 | 4 (1D device) | Monolithic | TE/TM | NA | 8 |

The excellent performance of our devices is attributed to the exceptionally large Faraday rotation and low loss of the Ce:YIG thin films. The Faraday rotation of the film can be inferred using Eqs. 1 (Supplementary Section 7) to be -2960 deg/cm, for Ce:YIG deposited on the Si TM device. This values is significantly higher than previously reported Ce:YIG thin films deposited by pulsed laser deposition (PLD)[10], and benefits from judicious control of the deposition oxygen partial pressure to drive higher $Ce^{3+}/Ce^{4+}$ ratios (Supplementary Fig. S2). The material and device losses are parameterized in Supplementary Section 7. Taking the TM isolator as an example, the total insertion loss of 5-6 dB mainly includes a 0.7 dB excess loss from each of the 3 dB MMI couplers, a propagation loss of 2.2-3.2 dB from the magneto-optical waveguides covered with garnet, a coupling loss of 0.25 dB/junction at the junctions between waveguide with and without garnet, and a propagation loss of 0.36 dB from the silicon waveguides not covered by garnet. Therefore the total insertion loss can be further reduced by optimizing the coupler and junction designs, for example by using low loss broadband adiabatic couplers[30] instead of MMIs, and using taper designs to minimize the waveguide junction losses. On the other hand, by further improving the Ce:YIG and YIG figure of merit[17], the material loss can be improved. Reducing the YIG seed layer thickness or using a top seed layer can also lead to a much higher device FOM by increasing coupling of light from the waveguide into the Ce:YIG layer[18,20]. Therefore a broadband monolithic isolator device with < 3 dB insertion loss is well within the reach of state-of-the art silicon photonic device technologies.

In summary, we experimentally demonstrated monolithically integrated broadband optical isolators for both TE and TM polarizations on silicon and on SiN platforms for the first time. The devices are based on high quality garnet polycrystalline thin films which we have grown on Si with properties comparable to those of epitaxial thin films, as well as garnet films grown on the sidewalls of Si and SiN waveguides for fully-TE isolator devices. These materials developments enabled the first demonstration of high performance monolithically integrated broadband optical isolators on Si and SiN for TE as well as TM modes, and a fully-TE ring resonator device on SiN. The devices feature high isolation ratio, low insertion loss, broadband operation, and small footprint. The work represents an important step towards practical implementation of monolithic isolator integration on-chip with performance approaching that of traditional bulk isolators. The capability to integrate high-quality magneto-optical thin films with photonic waveguides, validated through this work, also paves the path to experimental demonstration of several theoretically proposed magneto-optical photonic crystal structures[33,34] as well as to isolators, magneto-optical modulators and phase shifters, and topological photonic devices based on time-reversal symmetry breaking using magneto-optical materials[2].

## Methods

**Device fabrication:** The TM and TE isolators were fabricated on SOI and SiN platforms. For SOI devices, MTI Corp. SOI wafers with 220 nm device layer and 2 µm buried oxide were first cleaned in Piranha solutions for 10 minutes to remove any organic contaminations. A 4% HSQ resist (XR-1541, Dow Corning) was spun onto the wafer with thickness of ~100 nm and then exposed on an Elionix ELS-F125 electron beam lithography (EBL) system with a beam current of 8 nA. The resist was then developed in 25% tetramethylammonium hydroxide (TMAH) for 3 minutes to reveal device pattern. Reactive ion etch (RIE) with $Cl_2$ gases was subsequently utilized to transfer the pattern into the SOI wafer in a PlasmaTherm Etcher. Similarly, silicon nitride devices started from piranha cleaning a silicon wafer with 3µm thermal oxide, and then a 400 nm SiN device layer was deposited onto the wafer by LPCVD. Device patterning was performed with ZEP520A resist in the EBL system and the resist was developed in ZED-N50 for 1 minute. RIE was conducted in the same etcher with a gas mixture of $CHF_3$ and $CF_4$. Starting from this point, the processes described below were identical for SOI and SiN devices. A layer of FOX-25 (Dow Corning flowable oxide) was then spun onto the wafer with thickness of 400 nm followed by rapid thermal annealing (RTA) at 800 ºC for 5 minutes to form a planarized top $SiO_2$ cladding. An additional 250 nm PECVD silicon oxide was further deposited onto the wafer to completely isolate the optical mode from interacting with Ce:YIG deposited in next steps. Next, a second EBL process using a positive resist (ZEP520A) was carried out to pattern the window regions. Finally, for TM devices, buffered oxide etch (BOE) was used to expose the silicon waveguide surface. For TE devices, RIE using a gas mixture of $CHF_3$ and Ar ambient was applied to etch down silicon oxide top cladding and exposed one sidewall of the silicon waveguides. A piranha solution was used to clean the samples to remove any fluorinated polymer generated during the etching process. The as-fabricated devices were loaded into the PLD chamber for magneto-optical thin film deposition. Thin film deposition utilized a KrF excimer laser source which operates at 248 nm and at a repetition rate of 10 Hz. The fluence of the laser was determined to be 2.5 $J/cm^2$. The distance between target and substrate was fixed at 5.5 cm. 50 nm thick YIG thin films were first deposited onto the substrate at 450 ºC and then rapid thermal annealed (RTA) at 900 ºC for 5 minutes for fully crystallization. Finally, 100 nm thick Ce:YIG thin films were deposited at 650 ºC onto the devices.

**Device characterization:** The optical isolators were characterized on a fiber butt coupled waveguide test station. A LUNA Technology OVA 5000 was used to emit laser light from 1520 nm to 1610 nm. The transmitted light was then acquired by the OVA to analyze polarization dependent transmission spectra. In a different set-up, a free-space polarization control bench was used to obtain TE or TM polarized light before coupling to a polarization maintaining (PM) fiber. The linear polarized light was then butt coupled to the device for transmittance measurements with a lens tipped PM fiber. The testing methods were detailed in Supplementary materials. All devices were tested at least 3 times by reversing light propagation directions. The samples were maintained at room temperature with $\pm$ 0.2 ºC accuracy during the test.

## Acknowledgements

The authors are grateful to the support by National Natural Science Foundation of China (61475031, 51522204), Ministry of Science and Technology of China MOST (2016YFA0300802), and the US National Science Foundation (Award 1607865). Shared experimental facilities of CMSE, an NSF MRSEC under award DMR1419807, were used.

## Conflict of interests

The authors declare that they have no conflict of interest.

## Author contributions

C. A. R., J. H., J. B. and L. B. conceived and designed the project. Y. Z. and Q. D. fabricated the isolator devices. C. W. and T. F. deposited and characterized the magneto-optical thin films. S. L. designed the isolator devices. D. H., P. P. and L. J. D. contributed to the data analysis. All authors discussed the results and contributed to writing the manuscript.